\documentclass{jpsj-suppl}
\usepackage{txfonts} 

\title{Electromagnetic and Weak Nuclear Structure Functions $F_{1,2}(x,Q^2)$ in the Intermediate Region of $Q^2$ }
\author{H. Haider$^{1}$, F. Zaidi$^{1}$, M. Sajjad Athar$^{1}$, S. K. Singh$^{1}$, I. Ruiz Simo$^{2}$}
\inst{$^1$Department of Physics, Aligarh Muslim University, Aligarh - 202 002, India, $^2$Departamento de F\'{\i}sica At\'omica, Molecular y Nuclear,
and Instituto de F\'{\i}sica Te\'orica y Computacional Carlos I,
Universidad de Granada, Granada 18071, Spain}
\email{huma.haider8@gmail.com}

\recdate{January 19, 2016}

\abst{We have studied nuclear structure functions $F_{1A}(x,Q^2)$ and $F_{2A}(x,Q^2)$ for electromagnetic and weak processes 
 in the region of $1~GeV^2~<~Q^2~<8~GeV^2$. The nuclear medium effects arising due to Fermi motion, binding energy, nucleon correlations, mesonic contributions and 
shadowing effects are taken into account using a many body field theoretical approach. The calculations are performed in a local density approximation 
using a relativistic nucleon spectral function. The results are compared with the available experimental data. Implications of nuclear medium effects 
on the validity of Callan-Gross relation are also discussed.}
\kword{nuclear medium effect, structure functions, deep inelastic scattering, Callan-Gross relation}
\begin{document}
\maketitle
\section{Introduction}
 Recently a better understanding of nuclear medium effects in the deep inelastic scattering region both
 in electromagnetic(EM) and weak(Weak) interaction induced processes has been emphasized due to the fact 
 that experiments are being performed using electron beam at JLab~\cite{Mamyan:2012th} and neutrino/antineutrino beam at
 the Fermi Lab. The experiments are being performed using
several nuclear targets. The differential and total scattering cross sections 
are expressed in terms of $F_{1A}^{EM}(x,Q^2)$ 
and $F_{2A}^{EM} (x,Q^2)$ structure functions for electromagnetic 
processes and for weak interaction induced processes in 
terms of $F_{1A}^{Weak}(x,Q^2)$ , $F_{2A}^{Weak} (x,Q^2)$ and $F_{3A}^{Weak}(x,Q^2)$ structure functions. We have studied nuclear
 medium effects arising due to Fermi motion, binding energy, nucleon correlations, mesonic 
 contributions and shadowing effects in these structure functions, in a many body field theoretical
 approach and the calculations are performed
 in a local density approximation using a relativistic nucleon spectral function. The details of the present formalism
 are given in Ref.\cite{Haider:2015vea}. In this paper, we have compared  $F_{1A}^{EM}(x,Q^2)$ vs $F_{1A}^{Weak}(x,Q^2)$, and 
 $F_{2A}^{EM}(x,Q^2)$ vs $F_{2A}^{Weak}(x,Q^2)$ structure functions. The results for the ratio 
 $\frac{2xF_{1A}(x,Q^2)}{F_{2A}(x,Q^2)}$ with nuclear medium effects in carbon has also been presented. The results are also compared 
 with some of the available experimental data. For completeness, we are presenting the formalism 
 in brief.
\section{Formalism}
The double differential cross section for the reaction $\nu_l/\bar\nu_l(k) + N(p) \rightarrow l^-/l^+(k^\prime) + X(p^\prime),~l=~e^-,~\mu^-,$
 in the Lab frame is written as
\begin{equation} 	\label{dif_cross_nucleus}
\frac{d^2 \sigma_{\nu,\bar\nu}^N}{d \Omega' d E'} 
= \frac{G_F^2}{(2\pi)^2} \; \frac{|{\bf k}^\prime|}{|{\bf k}|} \;
\left(\frac{m_W^2}{q^2-m_W^2}\right)^2
L^{\alpha \beta}_{\nu, \bar\nu}
\; W_{\alpha \beta}^N\,,
\end{equation}
where $L^{\alpha \beta}_{\nu, \bar\nu}=k^{\alpha}k'^{\beta}+k^{\beta}k'^{\alpha}
-k.k^\prime g^{\alpha \beta} \pm i \epsilon^{\alpha \beta \rho \sigma} k_{\rho} k'_{\sigma}$ is the leptonic tensor with (+ve)-ve sign for (anti)neutrino, and 
$W_{\alpha \beta}^N$ is the nucleon hadronic tensor expressed in terms of nucleon structure functions $W_i^N;~i=1-3$,
\begin{eqnarray}\label{had_ten}
W^N_{\alpha \beta} =
\left( \frac{q_{\alpha} q_{\beta}}{q^2} - g_{\alpha \beta} \right) \;
W_1^N
+ \frac{1}{M_N^2}\left( p_{\alpha} - \frac{p . q}{q^2} \; q_{\alpha} \right)
\left( p_{\beta} - \frac{p . q}{q^2} \; q_{\beta} \right)
W_2^N -\frac{i}{2 M_N^2} \epsilon_{\alpha \beta \rho \sigma} p^{\rho} q^{\sigma}
W_3^N.
\end{eqnarray}
In a nuclear medium the expression for the cross section is written as
\begin{equation} 	\label{dif_crossA}
\frac{d^2 \sigma_{\nu,\bar\nu}^A}{d \Omega' d E'} 
= \frac{{G_F^2}}{(2\pi)^2} \; \frac{|{\bf k}^\prime|}{|{\bf k}|} \;
\left(\frac{m_W^2}{q^2-m_W^2}\right)^2
L^{\alpha \beta}_{\nu, \bar\nu}
\; W_{\alpha \beta}^{A}\,,
\end{equation}
$W^A_{\alpha \beta}$ is the nuclear hadronic tensor defined in terms of nuclear structure functions $W_i^A;~i=1-3$,
\begin{eqnarray}\label{had_tenA}
W^A_{\alpha \beta} =
\left( \frac{q_{\alpha} q_{\beta}}{q^2} - g_{\alpha \beta} \right) \;
W_1^A
+ \frac{1}{M_A^2}\left( p_{\alpha} - \frac{p . q}{q^2} \; q_{\alpha} \right)
\left( p_{\beta} - \frac{p . q}{q^2} \; q_{\beta} \right)
W_2^A - \frac{i}{2 M_A^2} \epsilon_{\alpha \beta \rho \sigma} p^{\rho} q^{\sigma}
W_3^A
\end{eqnarray}
where $M_A$ is mass of the target nucleus. The neutrino-nucleus cross sections are written 
 in terms of neutrino self energy $\Sigma(k)$ in the nuclear medium, 
 the expression for which is obtained as~\cite{Haider:2015vea}:
\begin{eqnarray}\label{Sigma}
\Sigma (k) = \frac{-iG_F}{\sqrt{2}}\frac{4}{m_\nu}
\int \frac{d^4 k^\prime}{(2 \pi)^4} \frac{1}{{k^\prime}^2-m_l^2+i\epsilon} 
\left(\frac{m_W}{q^2-m_W^2}\right)^2 \; L_{\alpha \beta} ~~ \Pi^{\alpha \beta} (q)\,.
\end{eqnarray}
$\Pi^{\alpha \beta} (q)$  is the W-boson self energy in the nuclear medium which is given in terms of nucleon ($G$) and meson ($D_j$)  propagators
\begin{eqnarray}\label{Self_1}
-i\Pi^{\alpha \beta} (q) &=&
(-) \; \int \frac{d^4 p}{(2 \pi)^4} iG(p) \;  \;
\sum_X \; \sum_{s_p, s_i} \prod^n_{i = 1}
\int \frac{d^4 p'_i}{(2 \pi)^4}
\prod_l i G_l (p'_l) \prod_j \; i D_j (p'_j) 
\nonumber\\
&& \times \left( \frac{-G_F m_W^2}{\sqrt{2}} \right)
\langle X | J^{\alpha} | N \rangle \langle X | J^{\beta} | N \rangle^*
(2 \pi)^4 \delta^4 (q + p - \Sigma^n_{i = 1} p'_i).~~~~~~~~~
\end{eqnarray}
Using the expression of W self-energy and neutrino self-energy in the expression of cross section 
one obtains~\cite{Haider:2015vea}:
\begin{equation}	\label{conv_WA}
W^A_{\alpha \beta} = 4 \int \, d^3 r \, \int \frac{d^3 p}{(2 \pi)^3} \, 
  \int^{\mu}_{- \infty} d p^0 \frac{M}{E ({\bf p})} S_h (p^0, {\bf p}, \rho(r))
W^N_{\alpha \beta} (p, q), 
\end{equation}
where $\mu$ is the chemical potential. The hole spectral function $S_h$ takes care of Fermi momentum, Pauli blocking, binding energy and
nucleon correlations~\cite{Marco:1995vb}. $W_i^N(x, Q^2 )$ and $W_i^A(x, Q^2 )$
are respectively redefined in terms of the dimensionless structure functions $F_i^N(x, Q^2 )$ and $F_i^A(x, Q^2 )$ through
\begin{eqnarray}
 M W_1^N(\nu,Q^2) &=& F_1^N(x,Q^2);~~~~~~~~M_A W_1^A(\nu,Q^2) = F_1^A(x,Q^2)\nonumber\\
 \nu W_2^N(\nu,Q^2) &=& F_2^N(x,Q^2);~~~~~~~~\nu_A W_2^A(\nu,Q^2) = F_2^A(x,Q^2)\nonumber\\
 \nu W_3^N(\nu,Q^2) &=& F_3^N(x,Q^2);~~~~~~~~\nu_A W_3^A(\nu,Q^2) = F_3^A(x,Q^2)\nonumber
\end{eqnarray}
The nucleon structure functions are expressed in terms of  parton
distribution functions(PDFs). For the numerical calculations, we have 
used CTEQ6.6~\cite{cteq} nucleon PDFs. The evaluations are 
performed both at the leading order(LO) and next-to-leading order(NLO). 
 For electromagnetic interaction, we follow the same procedure, formalism for which is given in accompanying paper by Zaidi et al.~\cite{Farhana} in this proceeding.
 
Expressing  $W^N_{\alpha \beta}$ and $W^A_{\alpha \beta}$,
in terms of $F^N_i$ and $F^A_i$ (i=1,2), we get~\cite{Haider:2015vea}  
\begin{eqnarray}	\label{conv_WA1}
F_{_{1~A}}^{EM/Weak}(x_A, Q^2) &=& 2\sum_{\tau=p,n} AM \int \, d^3 r \, \int \frac{d^3 p}{(2 \pi)^3} \, 
\frac{M}{E ({\bf p})} \, \int^{\mu}_{- \infty} d p_0 S_h^\tau (p_0, {\bf p}, \rho^\tau(r)) \times \nonumber\\
&&\left[\frac{F_{1}^{EM/Weak,\tau}(x_N, Q^2)}{M}+ \frac{{p_x}^2}{M^2} \frac{F_{2}^{EM/Weak,\tau}(x_N, Q^2)}{\nu}\right],
\end{eqnarray}

 \begin{eqnarray} 
F_{_{2~A}}^{EM/Weak}(x_A,Q^2)  &=&  2\sum_{\tau=p,n} \int \, d^3 r \, \int \frac{d^3 p}{(2 \pi)^3} \, 
\frac{M}{E ({\bf p})} \, \int^{\mu}_{- \infty} d p_0 S_h^\tau (p_0, {\bf p}, \rho^\tau(r)) \times \nonumber \\
&&\left[\frac{Q^2}{q_z^2}\left( \frac{|{\bf p}|^2~-~p_{z}^2}{2M^2}\right) +  \frac{(p_0~-~p_z~\gamma)^2}{M^2} 
\left(\frac{p_z~Q^2}{(p_0~-~p_z~\gamma) q_0 q_z}~+~1\right)^2\right]~\times\nonumber\\
&& \left(\frac{M}{p_0~-~p_z~\gamma}\right) ~F_2^{EM/Weak,\tau}(x,Q^2),       
\end{eqnarray}

where $\gamma=\frac{q_z}{q_0}= \sqrt{1+\frac{4 M^2 x^2}{Q^2}}$.
The mesonic (pion and rho) cloud contributions are taken into account following the same procedure
 as for the nucleon, except the fact that now instead of nucleon spectral function, we
 use meson propagator to describe the meson propagation in the nuclear medium.
 For this also, we have used microscopic approach
by making use of the imaginary part of the meson propagators instead of spectral function,
and obtain $F_{2, \pi}^A (x)$~\cite{Marco:1995vb} as
\begin{eqnarray} \label{pion_f21}
F_{_{2, A, \pi}}^{EM/Weak}(x_\pi,Q^2)  &=&  -6 \int \, d^3 r \, \int \frac{d^4 p}{(2 \pi)^4} \, 
        \theta (p_0) ~\delta I m D (p) \;2m_\pi~\left(\frac{m_\pi}{p_0~-~p_z~\gamma}\right)\times \nonumber \\
&&\left[\frac{Q^2}{q_z^2}\left( \frac{|{\bf p}|^2-p_{z}^2}{2m_\pi^2}\right)  
+  \frac{(p_0-p_z\gamma)^2}{m_\pi^2} \left(\frac{p_z~Q^2}{(p_0-p_z\gamma) q_0 q_z}+1\right)^2\right] F_{_{2, \pi}}^{EM/Weak}(x_\pi)
\end{eqnarray}
where $x_\pi=-\frac{Q^2}{2p \cdot q}$ and $D(p)$ is the pion propagator in the nuclear medium given by 
 \begin{equation}\label{dpi}
D (p) = [ {p_0}^2 - {\bf {p}}\,^{2} - m^2_{\pi} - \Pi_{\pi} (p_0, {\bf p}) ]^{- 1}\,,
\end{equation}
where
\begin{equation}\label{pionSelfenergy}
\Pi_\pi=\frac{f^2/m_\pi^2 F^2(p){\bf {p}}\,^{2}\Pi^*}{1-f^2/m_\pi^2 V'_L\Pi^*}\,.
\end{equation}
Here, $F(p)=(\Lambda^2-m_\pi^2)/(\Lambda^2+{\bf {p}}\,^{2})$ is the $\pi NN$ form factor, $\Lambda$=1~$GeV$, $f=1.01$, $V'_L$ is
the longitudinal part of the spin-isospin interaction and $\Pi^*$ is the irreducible pion self energy containing
the contribution from particle - hole and delta - hole excitations.
 For the meson PDFs we have used the parameterization of Gluck
et al.~\cite{Gluck:1991ey}. Similarly, the contribution of the $\rho$-meson cloud to the structure function is
taken into account~\cite{Marco:1995vb}
\begin{eqnarray} \label{F2rho1}
F_{_{2, A, \rho}}^{EM/Weak}(x_\rho,Q^2)  &=& -12 \int \, d^3 r \, \int \frac{d^4 p}{(2 \pi)^4} \, 
        \theta (p_0) ~\delta I m D (p) \;2m_\rho~\left(\frac{m_\rho}{p_0~-~p_z~\gamma}\right) \times \nonumber \\
&&\left[\frac{Q^2}{q_z^2}\left( \frac{|{\bf p}|^2~-~p_{z}^2}{2m_\rho^2}\right)  
+  \frac{(p_0-p_z~\gamma)^2}{m_\rho^2} \left(\frac{p_z~Q^2}{(p_0-p_z~\gamma) q_0 q_z}+1\right)^2\right]F_{_{2, \rho}}^{EM/Weak}(x_\rho)~~~~
\end{eqnarray}
where $x_\rho=-\frac{Q^2}{2p \cdot q}$ and $D_{\rho} (p)$ is now the $\rho$-meson propagator in the medium given by:
\begin{equation}\label{dro}
D_{\rho} (p) = [ {p_0}^2 - {\bf{p}}\,^{2} - m^2_{\rho} - \Pi^*_{\rho} (p_0, {\bf p}) ]^{- 1}\,,
\end{equation}
where
\begin{equation}\label{rhoSelfenergy}
\Pi^*_\rho=\frac{f^2/m_\rho^2 C_\rho F_\rho^2(p){\bf{p}}\,^{2}\Pi^*}{1-f^2/m_\rho^2 V'_T\Pi^*}\,.
\end{equation}
In Eq.\ref{rhoSelfenergy}, $V'_T$ is the transverse part of the spin-isospin interaction, $C_\rho=3.94$, $F_\rho(p)=(\Lambda_\rho^2-m_\rho^2)/(\Lambda_\rho^2+{\bf{p}}\,^{2})$ is the $\rho NN$ form factor, 
$\Lambda_\rho$=1~$GeV$, $f=1.01$, and $\Pi^*$ is the irreducible rho self energy which contains 
the contribution of particle - hole and delta - hole excitations. 
We have used the same PDFs for the $\rho$ meson as for the pions~\cite{Gluck:1991ey}.

We have also included shadowing effect following 
the works of Kulagin and Petti~\cite{Petti2}. 
For the shadowing effect which is due to the constructive interference of amplitudes
arising from the multiple scattering of quarks inside the nucleus, Glauber-Gribov 
multiple scattering theory has been used.
Shadowing effect is a low x and low $Q^2$ phenomenon which becomes negligible for high x.
We label the results of spectral function(SF)
with meson cloud contribution and shadowing effect, as the results obtained with full prescription(Total).
 \begin{figure}[tbh]
\includegraphics[height=4.5 cm, width=7 cm]{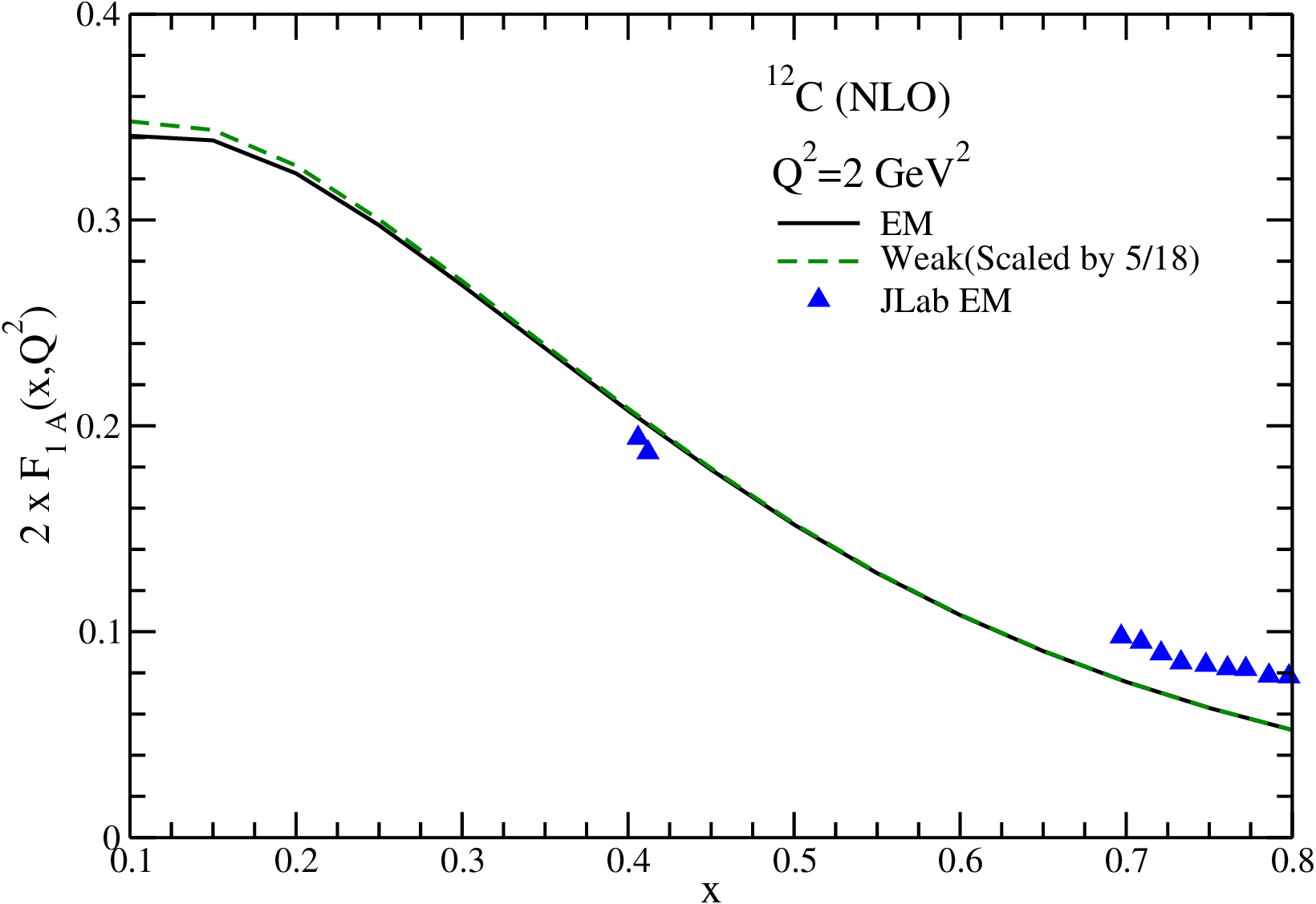}
\includegraphics[height=4.5 cm, width=7 cm]{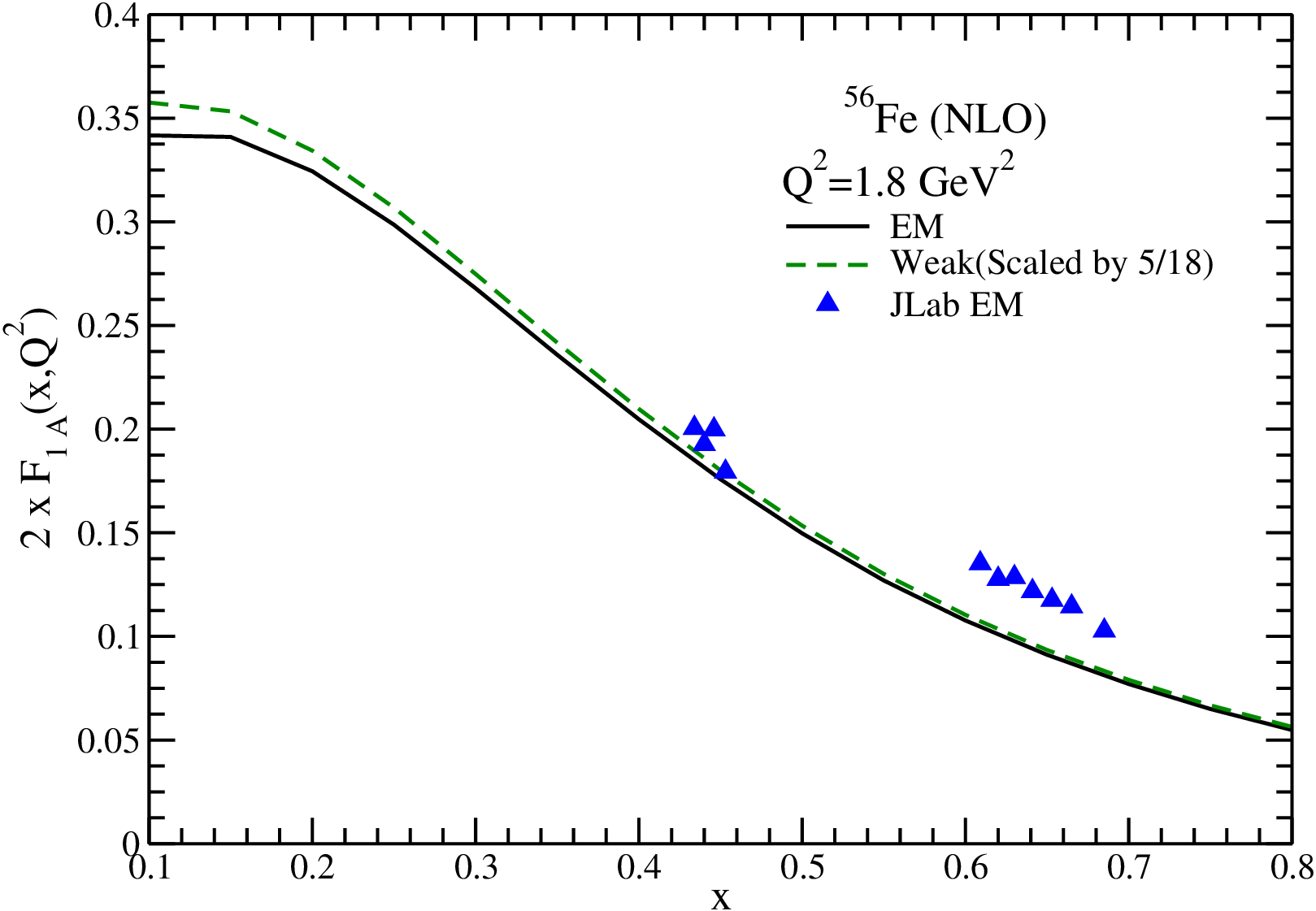}
\caption{Results for the $2 x F_{1~A}(x,Q^2)$ in $(A=)^{12}C$ and $^{56}Fe$ nuclei at NLO with full prescription. The experimental points are JLab data~\cite{Mamyan:2012th}.}
\label{f1}
\end{figure}
\section{Results}
In Fig.1, the results for $2xF_{1A}(x,Q^2)$ are shown for electromagnetic and weak 
 interactions in carbon and iron nuclei at a fixed $Q^2$ with full prescription for nuclear medium effect.
 The results for $2xF_{1A}^{EM}(x,Q^2)$ are compared with the JLab~\cite{Mamyan:2012th} data. We 
 found that the present results are consistent with the experimental data. 
From the figure, one may observe that at low x, EM structure function is slightly different 
 from weak structure function which is about $1-2\%$ for carbon. 
 This difference increases with the increase in mass number, for example, in iron it is $\sim 4\%$ at x=0.1.
 This difference is significant at low x 
and becomes almost negligible for high x.
This difference has also been found in the free nucleon case which shows a different distribution of
sea quarks for electromagnetic and weak interaction processes. Moreover, the difference in the case of 
nuclear target becomes a bit larger from the free nucleon case due to nuclear medium and nonisoscalarity effects.

 Similar is the observation for $F_{2A}(x,Q^2)$ in the electromagnetic as well as weak structure
 functions as may be observed in Fig.2.  
\begin{figure}[tbh]
\includegraphics[height=4.5 cm, width=7 cm]{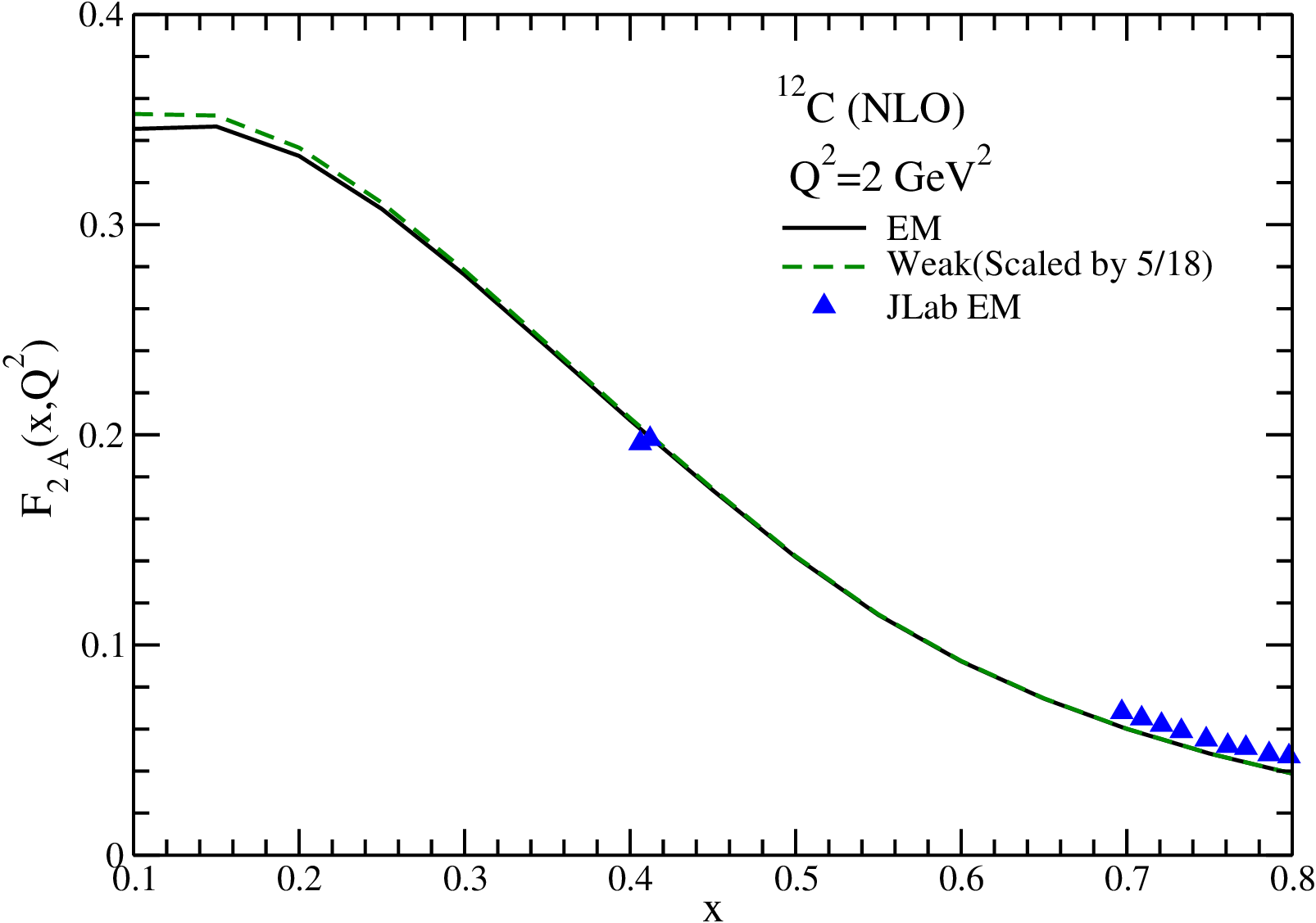}
\includegraphics[height=4.5 cm, width=7 cm]{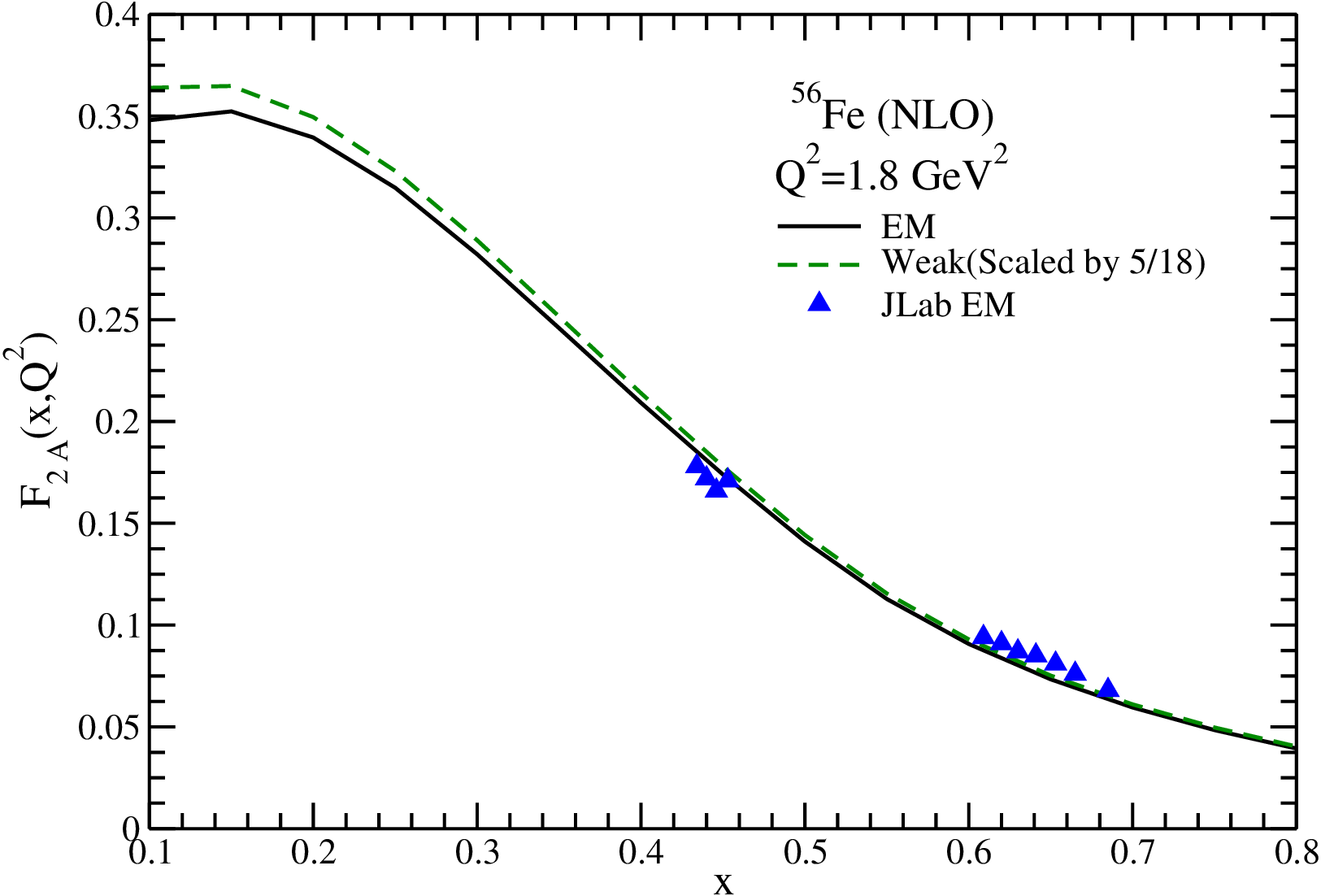}
\caption{Results for $ F_{2~A}(x,Q^2)$ in $(A=)^{12}C$ and $^{56}Fe$ nuclei at NLO with full prescription. 
The experimental points are JLab data~\cite{Mamyan:2012th}.}
\label{f2}
\end{figure}
In Fig.3 (left panel), we have shown the effect of mesonic cloud contribution 
and shadowing effect on the electromagnetic structure functions $F_{1A}(x,Q^2)$ and $F_{2A}(x,Q^2)$. For this,
we are presenting the results using the expression 
$r_i={\frac{F_i^{Modified}(x,Q^2)~-~F_i^{SF}(x,Q^2)}{F_i^{Modified}(x,Q^2)}}$, (i=1,2), for $^{56}Fe$ at $Q^2=5~GeV^2$,
 where $F_i^{SF}(x,Q^2)$ stands for the results obtained for the nuclear
structure functions using the spectral function(SF) 
 only while $F_i^{Modified}(x,Q^2)$ is the result obtained 
when  we include {\bf (i)} mesonic($\pi+\rho$) cloud contribution,
{\bf (ii)} mesonic($\pi+\rho$) cloud contribution and shadowing effect.
Mesonic cloud contribution is effective in the intermediate region of $x$ ($x \le 0.6$).
  The inclusion of shadowing effect hardly changes this ratio. 

     Furthermore, the effect of mesonic contributions 
is to increase $2 x F_{1A}(x,Q^2)$ and $F_{2A}(x,Q^2)$. The increase is larger at small 
$x$($x~~<~0.3$), for example $20\%$ at $x=0.2$, $12\%$ at $x=0.3$ and smaller at high $x$($x~~>~0.5$), for example
$\sim 2\%$ at $x=0.5$.
The increment in the results of $F_{2A}(x,Q^2)$ is more than in the results of 
$2 x F_{1A}(x,Q^2)$ over an entire range of $x$. 
 We have also studied this ratio in the case of
  weak interaction and found that the results almost overlap with the 
  results obtained for EM interaction. 
  
   \begin{figure}
\includegraphics[height=4.5 cm, width=7 cm]{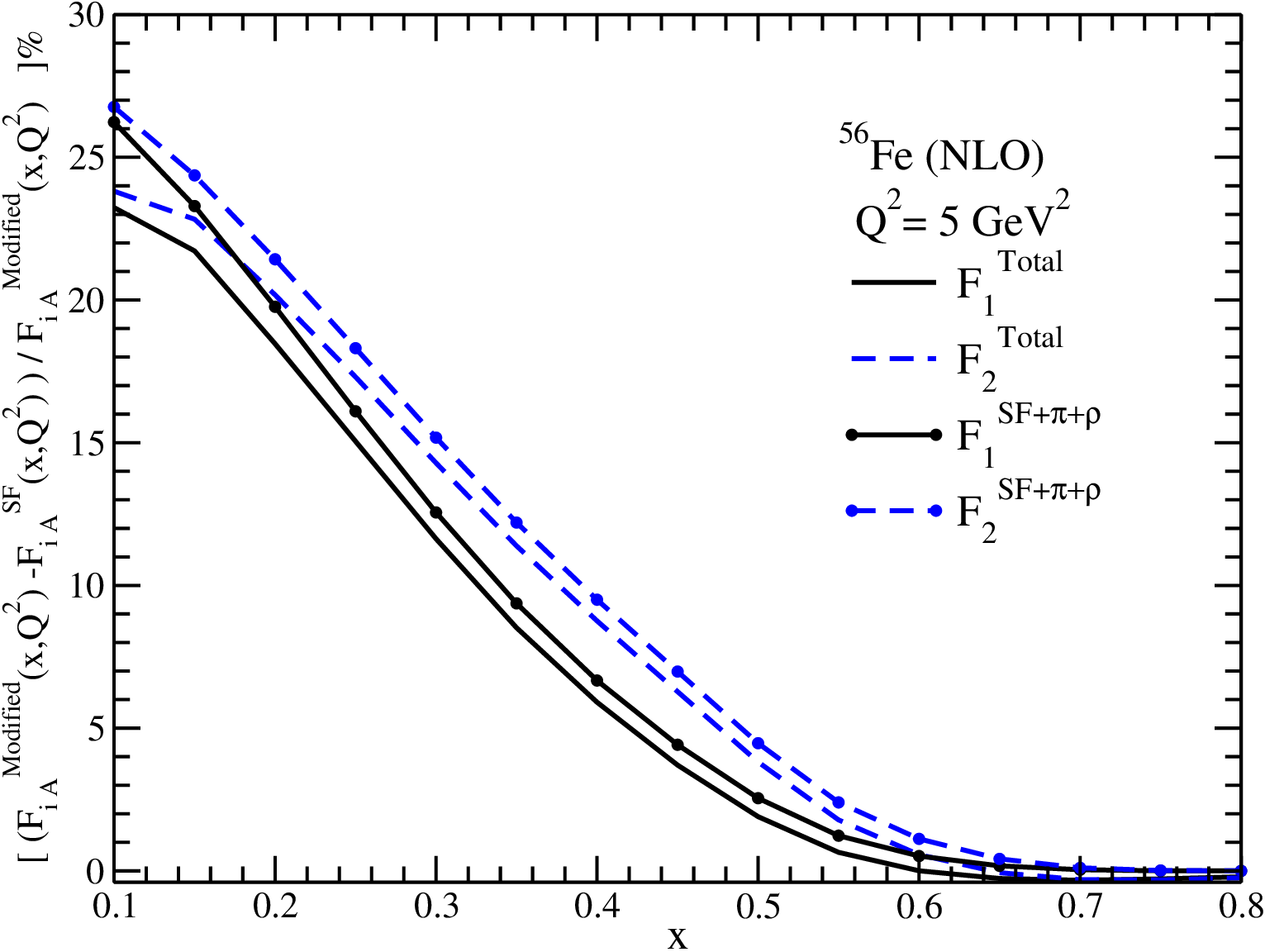}
\includegraphics[height=4.5 cm, width=7 cm]{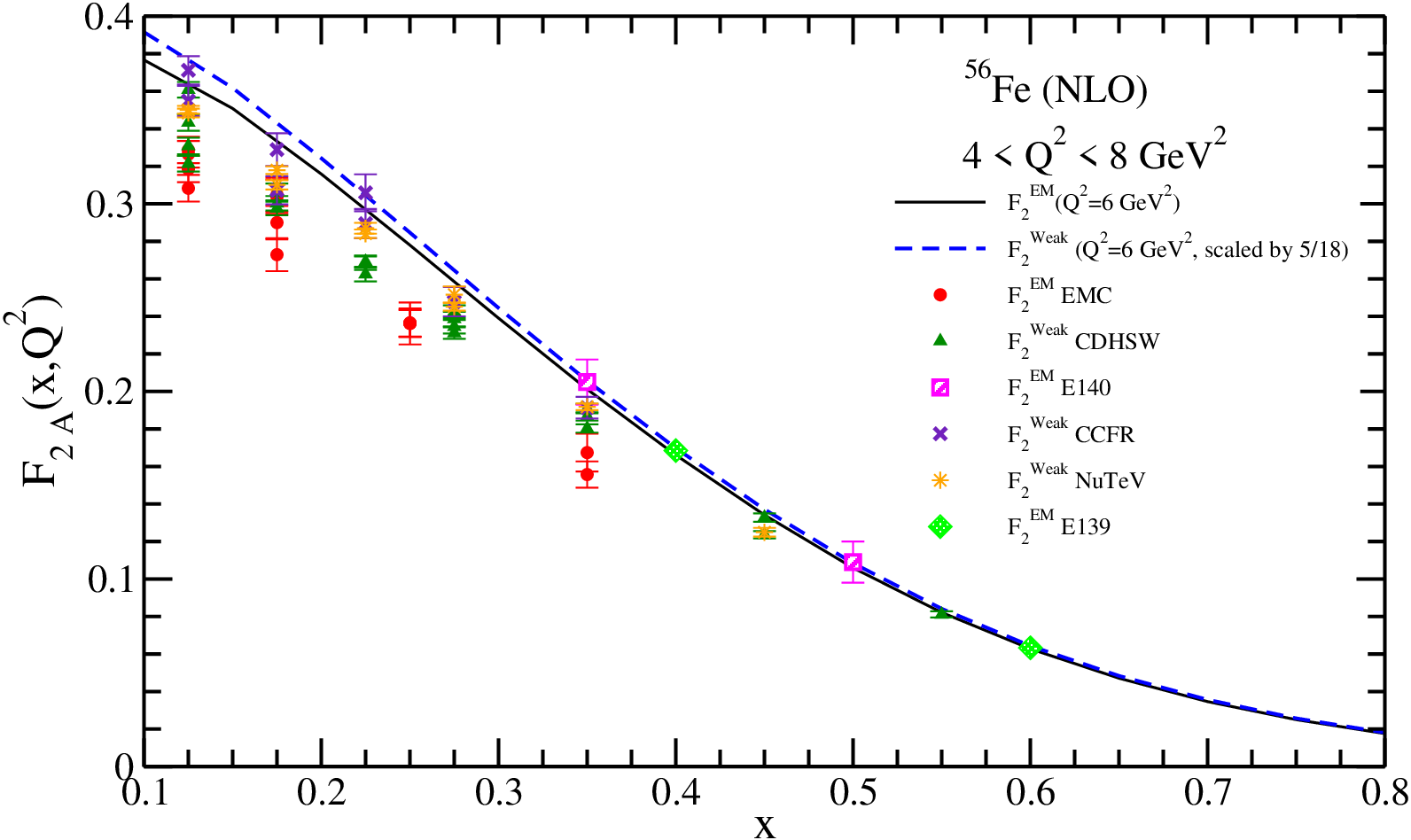}
\caption{{\bf Left panel:} $r_i$ vs $x$ in $(A=)^{56}Fe$ at $Q^2=5~GeV^2$ for EM interaction at NLO; 
{\bf Right panel:}
Results for $F_{2~A}(x,Q^2)$ in iron for EM and weak interactions at NLO 
with full prescription. The experimental points are data from
Refs.~\cite{Oltman:1992pq, Berge:1989hr, whitlow, Hen:2013oha, Dasu:1993vk, Aubert:1986yn, Tzanov:2005kr}. 
It may be noted that the experimental data points lying below 
the theoretical curves are from the older experiments which have measured 
$F_{2~A}^{EM}(x,Q^2)$~\cite{Aubert:1986yn} and $F_{2~A}^{Weak}(x,Q^2)$~\cite{Berge:1989hr}.}
\end{figure}
It may be observed that when shadowing effects are included there is net reduction in the result. 
This is because the shadowing and the mesonic effects tend to cancel each other specially
in the region of small $x$($x~~<~0.2$), where shadowing is important. 

In the right panel of Fig.3, we have presented the results for
 EM and weak structure functions $F_{2A}(x,Q^2)$ in iron at $Q^2=6~GeV^2$ and compared them 
 with some of the experimental data~\cite{Oltman:1992pq, Berge:1989hr, whitlow, Hen:2013oha, Dasu:1993vk, Aubert:1986yn, Tzanov:2005kr} 
 available for $F_{2A}^{EM}(x,Q^2)$ 
 and $F_{2A}^{Weak}(x,Q^2)$. One may observe that theoretically $F_{2A}^{EM}(x,Q^2)$ 
 lies below $F_{2A}^{Weak}(x,Q^2)$ over the entire region of $x$. 
 Furthermore, we observe that explicitly our model in agreement with the 
 experimental data of CCFR~\cite{Oltman:1992pq}, EMC139~\cite{whitlow, Hen:2013oha}, EMC140~\cite{Dasu:1993vk} and NuTeV~\cite{Tzanov:2005kr} 
 experiments. 

To quantify our results  in Fig.4, we present the results for 
the ratio of $\frac{2xF_{1A}(x,Q^2)}{F_{2A} (x,Q^2)}$ in carbon and compare it with the JLab data~\cite{Mamyan:2012th}. 
We have found that the ratio of $\frac{2xF_{1A}(x,Q^2)}{F_{2A} (x,Q^2)}$ is different than unity, i.e.
 the results presented here give a microscopic description of deviation 
from Callan-Gross relation($\frac{2xF_{1A}(x,Q^2)}{F_{2A} (x,Q^2)}=1$) due to nuclear medium effects. 
From the figure, one may observe 
that the ratio of EM and weak structure functions overlap each other.
 It would be interesting to make similar studies 
in the case of MINERvA experiment.
\begin{figure}
\begin{center}
\includegraphics[height=4.5 cm, width=10 cm]{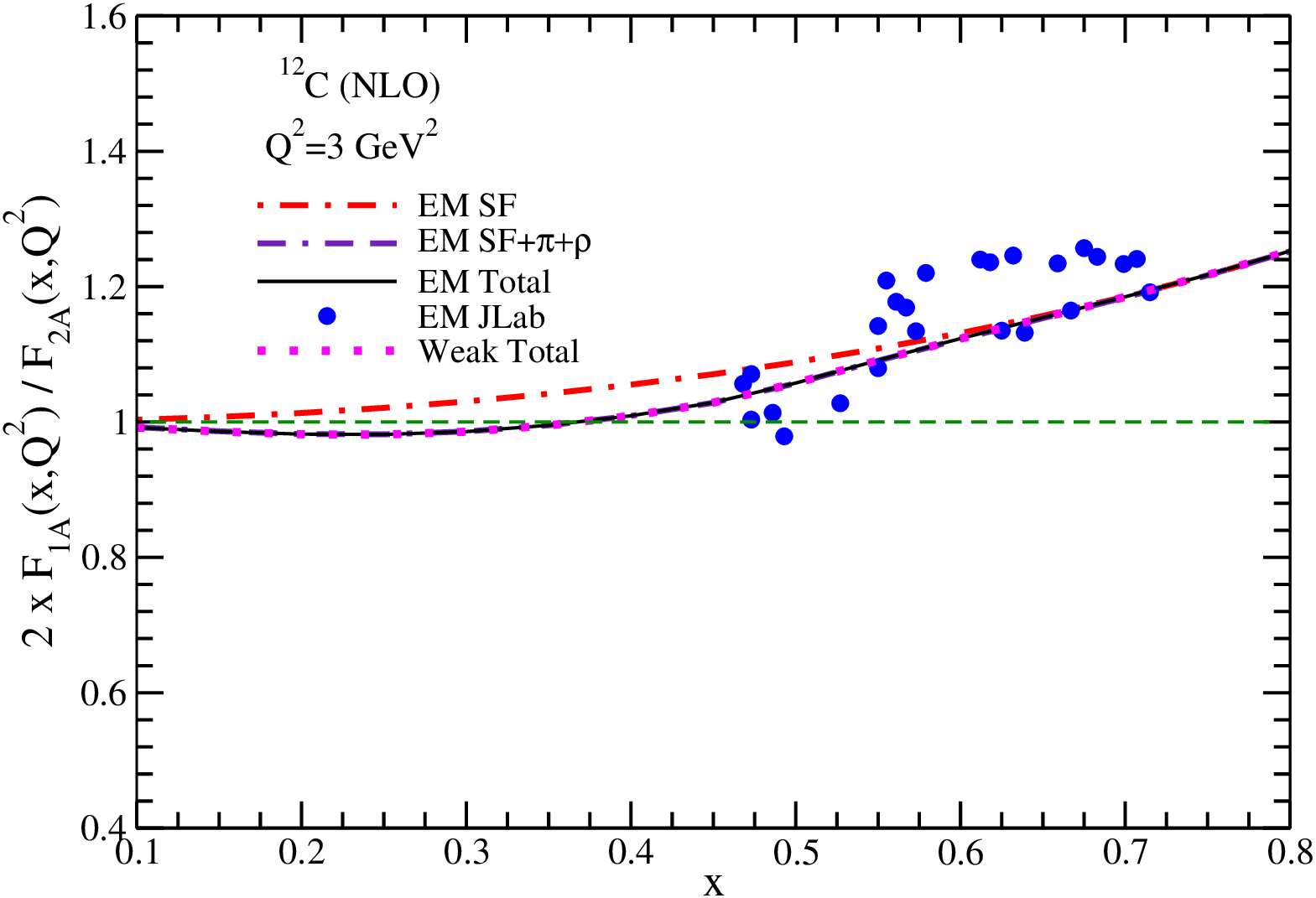}
\end{center}
\caption{Ratio of structure functions showing the violation of Callan-Gross relation 
at nuclear level. The experimental points are JLab data~\cite{Mamyan:2012th}. Dashed line corresponds 
to the Callan-Gross relation $\frac{2xF_{1A}(x,Q^2)}{F_{2A} (x,Q^2)}=1$. }
\end{figure}

\end{document}